\documentclass[prb,superscriptaddress,twocolumn,showpacs,preprintnumbers,amsmath,floatfix]{revtex4}
\usepackage{amssymb}
\usepackage{graphicx}

\begin{document}

\title{Orbital Selective Mott Transition Induced by Orbitals with Distinct Noninteracting Densities of States}
\author{Ze-Yi Song}
\affiliation{Shanghai Key Laboratory of Special Artificial Microstructure Materials and Technology, School of Physics Science and engineering, Tongji University, Shanghai 200092, P.R. China}
\author{Hunpyo Lee}
\affiliation{School of General Studies, Kangwon National University, 346 Jungang-ro, Samcheok-si, Kangwon-do, South Korea}
\author{Yu-Zhong Zhang}
\email[Corresponding author.]{Email: yzzhang@tongji.edu.cn}
\affiliation{Shanghai Key Laboratory of Special Artificial Microstructure Materials and Technology, School of Physics Science and engineering, Tongji University, Shanghai 200092, P.R. China}
\date{\today}

\begin{abstract}
\centerline{Abstract}
By applying dynamical mean-field theory in combination with exact diagonalization at zero temperature to a half-filled Hubbard model with two orbitals having distinct noninteracting densities of states, we show that an orbital selective Mott transition (OSMT) will take place even without crystal field splitting, differences in bandwidth and orbital degeneracy. We find that formation of local spin triplet states followed by a two-stage breakdown of the Kondo effect, rather than decoupling of charge degrees of freedom among different orbitals, is the underlying physics for the OSMT. The relevance of our findings to Ca$_{2-x}$Sr$_x$RuO$_4$ and the iron-based superconductors is discussed, and a decent candidate to detect such an origin for the OSMT is proposed.
\end{abstract}
\pacs{71.27.+a,71.30.+h,71.10.Hf,71.10.-w}

\maketitle

\section{Introduction\label{Introduction}}

Since unconventional superconductivity emerges in the proximity to the Mott metal-to-insulator transition~\cite{ImadaRMP1998, KurosakiPRL2005}, it is of great importance to understand the origin for the Mott physics. In single orbital cases, enhanced spin fluctuations by strong correlations win the competition against kinetic energy, leading to a complete localization of all electrons across the Mott transition~\cite{ImadaRMP1998}. However, such a simple origin can hardly be applied to various topical materials directly, such as iron-based superconductors~\cite{YiPRL2013,LiPRB2014}, Ca$_{2-x}$Sr$_x$RuO$_4$~\cite{NakatsujiPRL2000}, and $^3$He bilayer system~\cite{NeumannScience2007}, where localized and itinerant fermions coexist.

Orbital selective Mott transition (OSMT)~\cite{VojtaJLTP2010,GeorgesARCMP2013}, where part of itinerant electrons becomes localized due to the inclusion of orbital degrees of freedom, can account for the coexistence~\cite{YuPRL2013,BeachPRB2011,AnisimovEPJB2002}. Similar to the single orbital cases, it is of considerable interest to unveil the origin for the OSMT~\cite{KogaPRL2004,KnechtPRB2005,KogaPRB2005,AritaPRB2005,SongPRB2005,FerreroPRB2005,MediciPRB2005,BiermannPRL2005,LiebschPRL2005,InabaPRB2006,CostiPRL2007,BouadimPRL2009,JakobiPRB2009,SongEPJB2009,LeePRL2010,GregerPRL2013,GregerarXiv2013,WernerPRL2007,MediciPRL2009}. Till now, three origins have been identified. Those are: 1) orbitals with different bandwidths~\cite{AnisimovEPJB2002}; 2) large crystal field splitting~\cite{WernerPRL2007}; and 3) difference in orbital degeneracies~\cite{MediciPRL2009}.

However, controversies remain on the applicabilities of above origins to real materials. For example, 1) large difference in bandwidth is not present either in Ca$_{2-x}$Sr$_x$RuO$_4$~\cite{FangPRB2004,KoPRL2004} or in iron-based superconductors~\cite{MiyakeJPSJ2010}; 2) crystal field splitting may lead to a single metal-to-insulator transition in Ca$_{2-x}$Sr$_x$RuO$_4$~\cite{LiebschPRL2007}; and 3) the orbital degeneracy may be lifted if lattice distortion occurs in the iron-based superconductors~\cite{StewartRMP2007}. Therefore, a new origin which is independent of the bandwidths, the crystal field splitting, and the orbital degeneracies is required in order to account for the OSMT in general.

Recently, orbital selective phase transition (OSPT), a counterpart of the OSMT in magnetically ordered states, has been extensively investigated~\cite{LeePRB2010,LeePRB2011,ZhangPRB2012,YaoMPLB2013}. While it may account for the possible orbital selectivity in iron-based superconductors of low-symmetry magnetically ordered states, the OSPT can not be applied to the iron-based superconductors of high-symmetry paramagnetic phases and the paramagnetic Ca$_{2-x}$Sr$_x$RuO$_4$ at $0.2 \leq x \leq 0.5$. Furthermore, the underlying physical picture of the OSPT is believed to be the orbital decoupling due to the symmetry breaking.

Here, based on dynamical mean field theory (DMFT)~\cite{GeorgesRMP1996} with exact diagonalization (ED) as an impurity solver~\cite{CaffarelPRL1994,ZhangPRB2007}, we show that the OSMT can happen at half filling and $T=0$, provided different orbitals have distinct noninteracting densities of states (DOS), even if crystal field splitting, orbital degeneracy and difference in bandwidth are all absent. The phenomenon of each orbital with distinct noninteracting DOS is commonly present in layered materials with open $d$ shell. For example, in Ca$_{2-x}$Sr$_x$RuO$_4$, the DOS of d$_{yz/zx}$ orbital is quasi-one-dimensional like while that of d$_{xy}$ orbital quasi-two-dimensional like~\cite{AnisimovEPJB2002,LiebschPRL2007}. By analyzing the interorbital spin and charge fluctuations, we conclude that a two-stage breakdown of the Kondo screening~\cite{JayaprakashPRL1981} stabilized by the formation of local spin triplet state due to the interorbital spin fluctuations is the underlying scenario for the OSMT, rather than the physical picture proposed by de'Medici, {\it et al.} that each orbital behaves as a single band Hubbard model due to the decoupling of correlated orbitals by suppressing the interorbital charge fluctuations~\cite{MediciPRB2011,MediciPRL2014}.

The paper is organized as follows. In Sec.~\ref{MM} we present the model we studied and the details of our dynamical mean field theory calculations. In Sec.~\ref{Results} we present our results, including the densities of states, the self-energies in different phases, the renormalization factors, the correlation functions, as well as the phase diagram, and we discuss the underlying physics picture of the OSMT. Finally, in Sec.~\ref{CD} we discuss the relevance of our findings to the materials and present a summary.

\section{Model and Method\label{MM}}

In order to justify the above new origin for the OSMT, we use a minimal two-orbital Hubbard model on a two-dimensional lattice, defined as
\begin{eqnarray}
&H&=-\sum_{\langle ij\rangle \gamma \sigma} t^{\gamma}_{ij} c^{\dagger}_{i\gamma\sigma}c_{j\gamma\sigma} -\mu \sum_{i \gamma \sigma}n_{i\gamma\sigma}\label{eq:hamiltonian} \\
&+& U\sum_{i\gamma}n_{i\gamma\uparrow}n_{i\gamma\downarrow} +\Big(U'- J^z \Big)\sum_{i \sigma}n_{i a \sigma} n_{i b \sigma} \nonumber \\
&+& U'\sum_{i \sigma}n_{i a \sigma} n_{i b \bar{\sigma}} - J^{\pm} \sum_{i} \left[S^{+}_{i a}S^{-}_{i b} +S^{-}_{i a}S^{+}_{i b}\right] \nonumber \\
&-& J^{p} \sum_{i}\left[c^{\dagger}_{i a \uparrow}c^{\dagger}_{i a  \downarrow}c_{i b \uparrow}c_{i b \downarrow}
+c^{\dagger}_{i b \uparrow}c^{\dagger}_{i b \downarrow}c_{i a \uparrow}c_{i a \downarrow},\right] \nonumber
\end{eqnarray}
where $t^{\gamma}_{ij}=t^{\gamma}_x(t^{\gamma}_y)$ is the intra-orbital hopping integral between nearest neighbor sites along $x$ ($y$) direction denoted by
$\langle ij\rangle$ with orbital indices $\gamma=a,b$. Throughout the paper, $t^b_x=t$ is chosen as the unit of the energy. $U$, $U^{\prime }$ and
$J^z$, $J^{\pm}$, $J^{p}$ are the intra-orbital, inter-orbital Coulomb interaction and the Hund's
coupling divided into Ising term, spin flip term, paring hopping term, respectively. $c^{\dagger}_{i\gamma\sigma}$
($c_{i\gamma\sigma}$) creates (annihilates) an electron in orbital
$\gamma$ of site $i$ with spin $\sigma$. $n_{i\gamma\sigma}=c^{\dagger}_{i\gamma\sigma}c_{i\gamma\sigma}$ is the occupation operator, while $S^{+}_{i\gamma}=c^{\dagger}_{i\gamma\uparrow}c_{i\gamma\downarrow}$ the spin operator. We are interested in two cases: 1) the isotropic case where $J^{\pm}=J^{p}=J^{z}$, and 2) the anisotropic case where $J^{\pm}=J^{p}=0$. Both cases satisfy the condition of $U=U^{\prime }+2J^{z}$. We consider the chemical potential $\mu=U/2+U'-J^{z}/2$, where both bands are half filled.

We investigate the ground state properties of model~(\ref{eq:hamiltonian}) in the paramagnetic state by combination of DMFT and ED~\cite{CaffarelPRL1994,ZhangPRB2007} where the two-orbital lattice model is mapped onto a two impurity Anderson model~\cite{GeorgesRMP1996} with each impurity coupled to $6$ discretized bath sites which are determined self-consistently through
\begin{eqnarray}
&&(g^{0}_{\gamma\sigma}\left( i\omega_n \right) )^{-1} - \Sigma_{\gamma\sigma} \left( i\omega_n \right) \\
&=&\left( \int \frac{d\epsilon \rho^{0}_{\gamma\sigma}\left( \epsilon \right) }{i\omega_n+\mu -\epsilon -\Sigma_{\gamma\sigma} \left( i\omega_n \right) } \right)^{-1}. \nonumber \label{selfconsistent}
\end{eqnarray}
Here, $i\omega_n$ is the Matsubara frequency, and $g^{0}_{\gamma\sigma}\left( i\omega_n \right) $ is the Weiss field where hybridization function of the impurity Anderson model is involved, while $\Sigma_{\gamma\sigma} \left( i\omega_n \right)$ is the local self-energy. The noninteracting DOS for each orbital is defined as
\begin{equation}
\rho^{0}_{\gamma\sigma}\left( \epsilon \right)=(1/2\pi)^2\int d\mathbf{k} \delta\left( \epsilon - E_{\gamma\sigma}(\mathbf{k}) \right)
\end{equation}
where the energy dispersion relations are chosen to be
\begin{equation}
E_{\gamma\sigma}(\mathbf{k})=-2t^{\gamma}_x cos(k_x)-2t^{\gamma}_y cos(k_y)
\end{equation}
with $t^{\gamma}_y/t^{\gamma}_x=\alpha<1$ for $\gamma=a$ and $1$ for $\gamma=b$, which simulate the difference between the DOS of d$_{yz/zx}$ orbital and that of d$_{xy}$ orbital in Ca$_{2-x}$Sr$_x$RuO$_4$~\cite{AnisimovEPJB2002,LiebschPRL2007}. Please note, in order to identify the role of each orbital having distinct noninteracting DOS, we neglect all the other one-particle terms, such as next-nearest-neighbor hoppings which may break the particle-hole symmetry and therefore induce an effective crystal field splitting. Moreover, we exclude the effect of difference in bandwidth by rescaling the bandwidths of the two orbitals to be the same, i.e., $W=8t$ for both orbitals. It is obvious that difference in orbital degeneracy is automatically precluded since the model we studied contains two orbitals only. Thus, if the OSMT occurs in model~(\ref{eq:hamiltonian}), the origin should be solely ascribed to the distinct noninteracting DOS of each orbital. In our calculations, we set an effective inverse temperature $\beta t=200$ which serves as a low-frequency cutoff.

\section{Results\label{Results}}

\begin{figure}[htbp]
\includegraphics[width=0.48\textwidth]{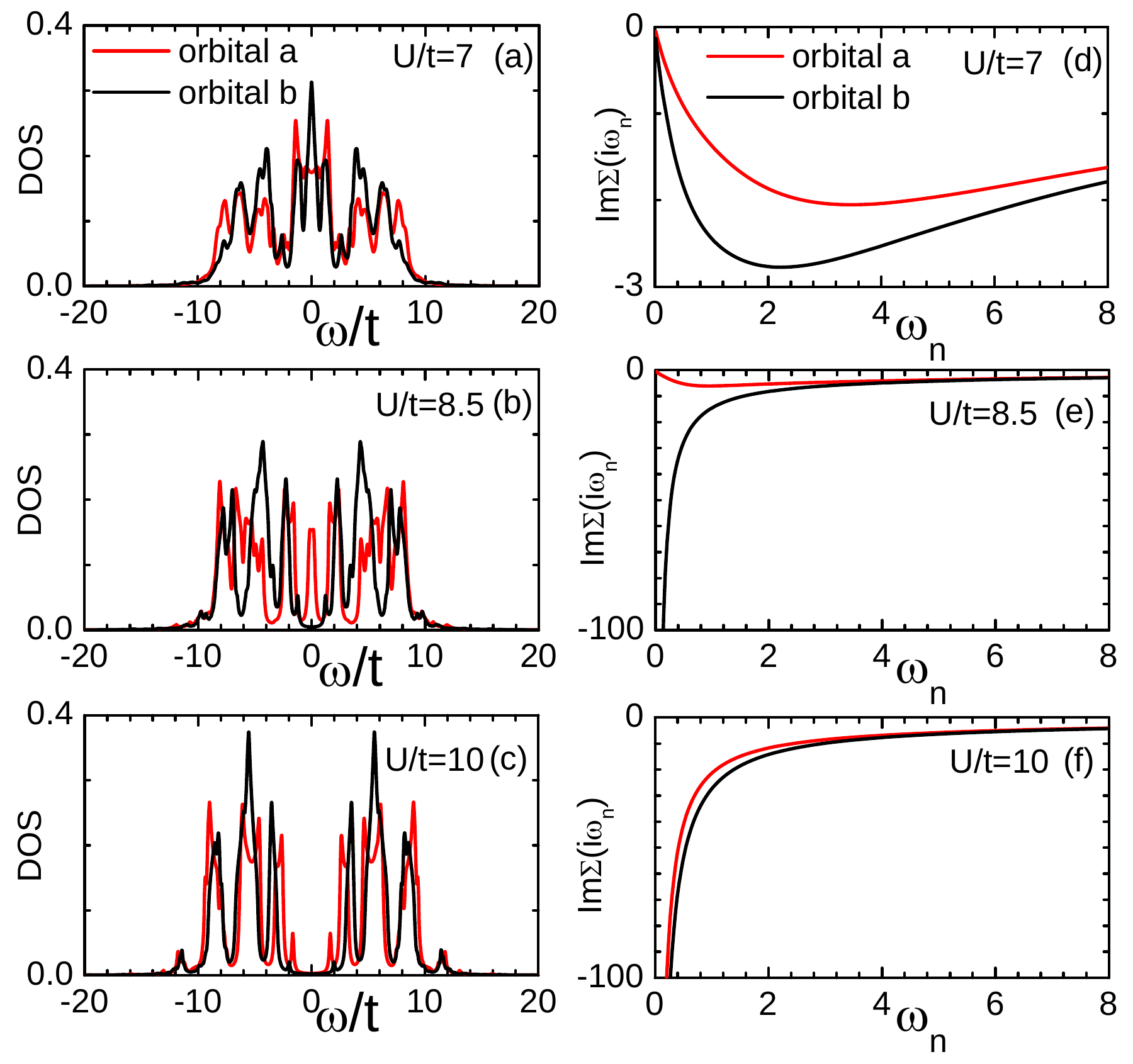}
\caption{(Color online) Orbitally resolved DOS ((a)-(c))and self-energies ((d)-(f)) at $J^{\pm}=J^{p}=J^{z}=U/4$ (isotropic case) and $\alpha=t^{a}_y/t^{a}_x=0.1$, $t^{b}_y/t^{b}_x=1$ for three different values of $U/t$. (a) and (d) is for $U/t=7$, indicating a metal, while (c) and (f) for $U/t=10$, denoting a Mott insulator. (b) and (e) represent for the OSMP where one orbital remains metallic while the other is Mott insulating.}
\label{Fig:one}
\end{figure}

\begin{figure}[htbp]
\includegraphics[width=0.48\textwidth]{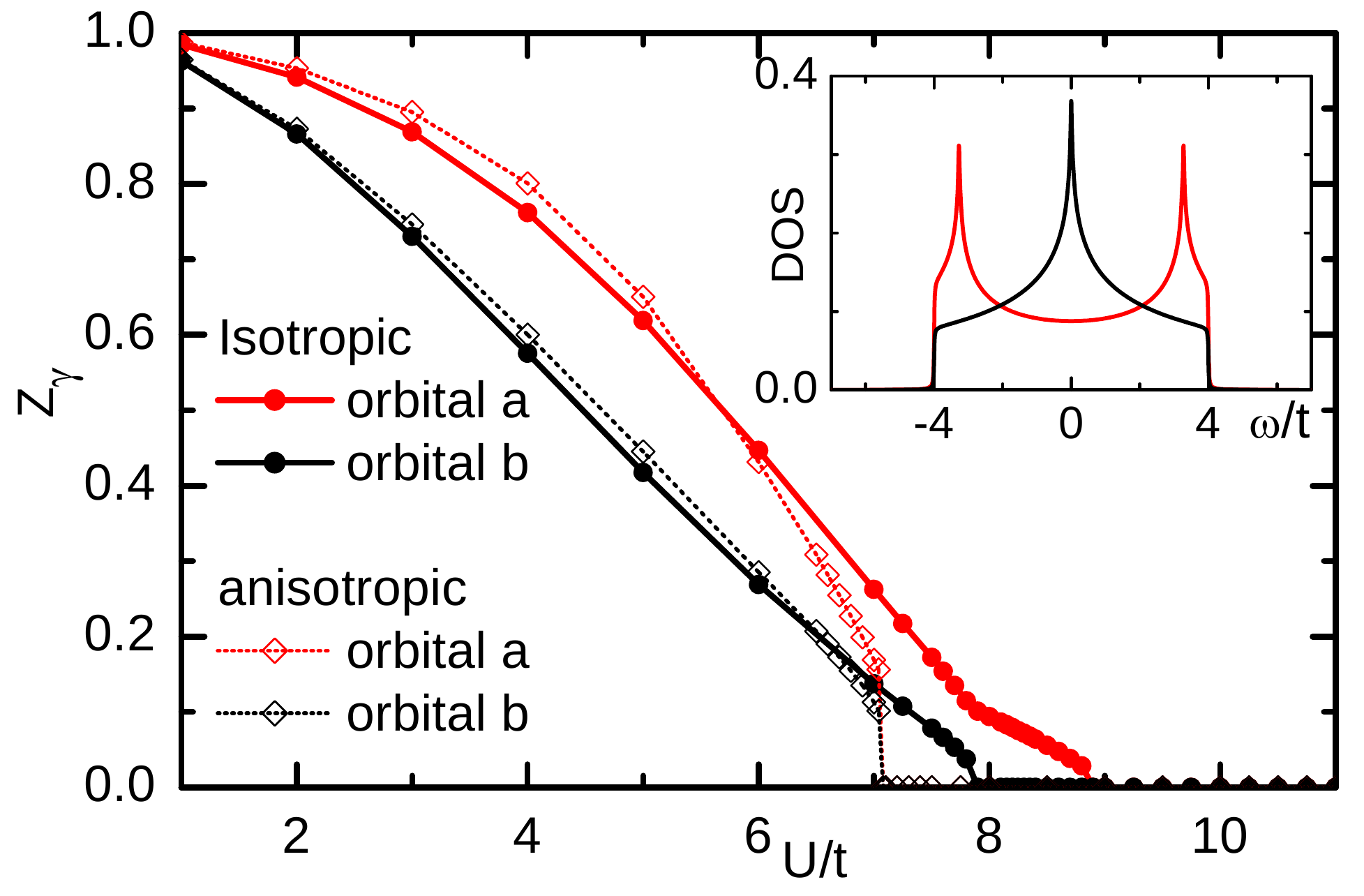}
\caption{(Color online) Renormalization factor $Z_{\gamma}$ as a function of $U/t$ in two cases: 1) the isotropic case where $J^{\pm}=J^{p}=J^{z}$ and 2) the anisotropic case where $J^{\pm}=J^{p}=0$. In both cases, we set $J^{z}=U/4$ and $\alpha=t^{a}_y/t^{a}_x=0.1$, $t^{b}_y/t^{b}_x=1$. The inset shows the noninteracting DOS for each orbital. It is found that an OSMP appears in the isotropic case while it disappears in anisotropic case.}
\label{Fig:two}
\end{figure}

Fig.~\ref{Fig:one} shows DOS and imaginary parts of self-energies for three different values of $U/t$ at $\alpha=t^{a}_y/t^{a}_x=0.1$, $t^{b}_y/t^{b}_x=1$ in the isotropic case with $J^{\pm}=J^{p}=J^{z}=U/4$. The noninteracting DOS of each orbital is shown in the inset of Fig.~\ref{Fig:two}. It is found in Fig.~\ref{Fig:one} (a) and (d) that at $U/t=7$, finite DOS is present at the Fermi level and the imaginary parts of self-energies approach zero at small Matsubara frequencies in both orbitals, indicating a metallic state. At large $U/t=10$, as seen in Fig.~\ref{Fig:one} (c) and (f), both orbitals exhibit zero DOS at the Fermi level and the imaginary parts of self-energies are both divergent at low frequencies, clearly revealing true Mott insulating behavior. At intermediate value of $U/t=8.5$, while both the DOS and the imaginary part of self-energy suggest that the metallic state remains in orbital a, orbital b exactly displays Mott insulating behavior with infinite scattering rate and zero DOS at the Fermi level, as shown in Fig.~\ref{Fig:one} (b) and (e). Since no difference between these two orbitals other than the noninteracting DOS is preserved in our model~(\ref{eq:hamiltonian}), we conclude that each orbital with distinct noninteracting DOS is the origin for the OSMT.

In Fig.~\ref{Fig:two}, we present the renormalization factor
\begin{equation}
Z_{\gamma}=\left( 1-\frac{\partial \text{Re} \Sigma_{\gamma}\left( \omega \right) }{\partial \omega }| _{\omega \rightarrow 0} \right)^{-1}
\end{equation}
as a function of $U/t$ at $\alpha=t^{a}_y/t^{a}_x=0.1$, $t^{b}_y/t^{b}_x=1$ in both the isotropic case with $J^{\pm}=J^{p}=J^{z}=U/4$ and the anisotropic case with $J^{z}=U/4,J^{\pm}=J^{p}=0$. We find that, in the isotropic case, the renormalization factor of orbital b first vanishes around $U/t=7.9$, and then that of orbital a disappears at larger value of $U/t=8.9$, indicating that consecutive Mott transitions happen from a metal to a Mott insulator intermediated by an OSMP. In contrast, in the anisotropic case, the renormalization factors in both orbitals simultaneously disappear around $U/t=7.06$ abruptly, implying that the OSMT can not survive without spin flip term. Please note, we have checked that the pair hopping term play a minor role in the OSMT.

\begin{figure}[htbp]
\includegraphics[width=0.48\textwidth]{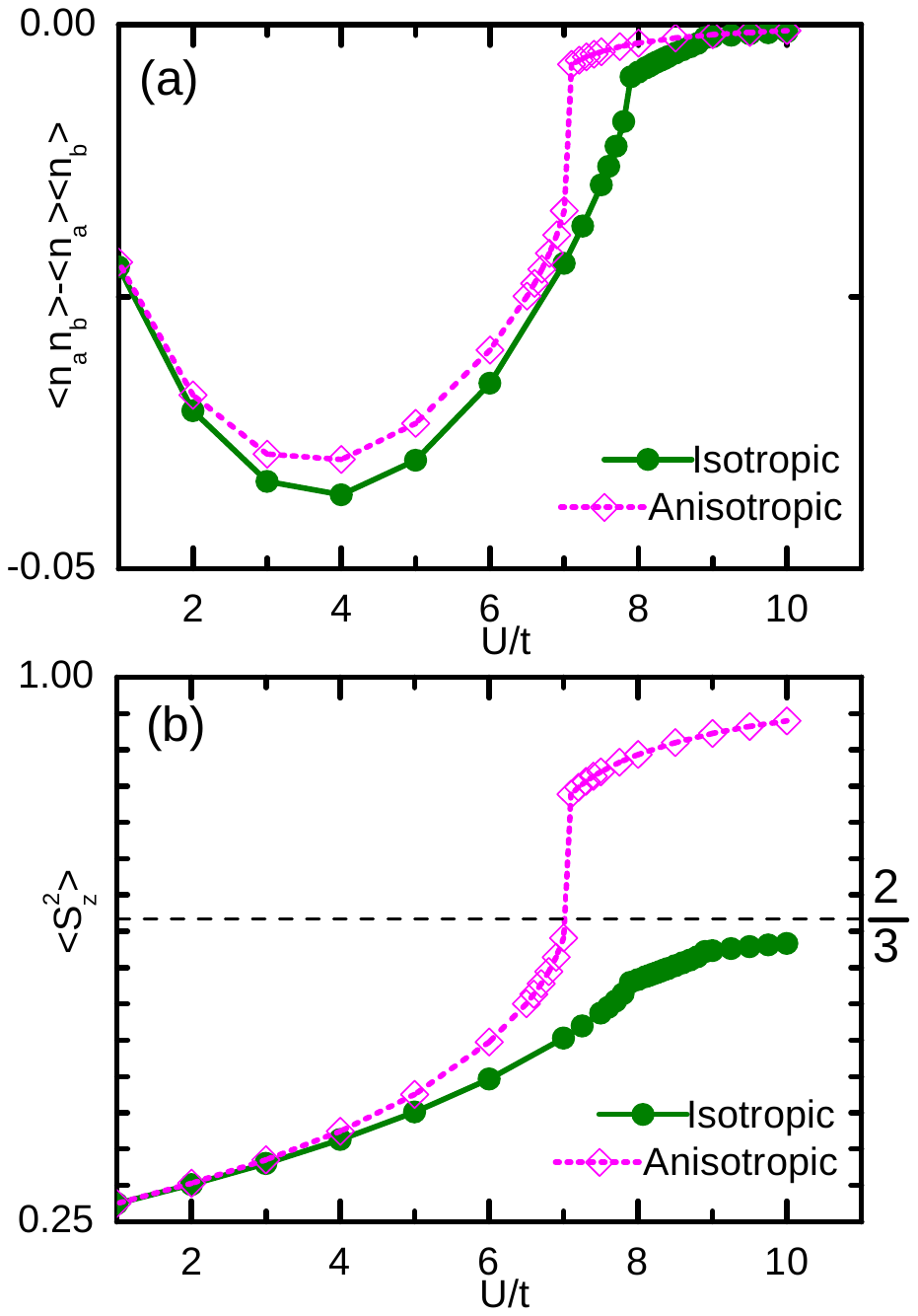}
\caption{(Color online) The interorbital charge fluctuations (a), squared total spin in $z$ direction (b) as a function of $U/t$, respectively, at $\alpha=t^{a}_y/t^{a}_x=0.1$ and $t^{b}_y/t^{b}_x=1$ for both isotropic case with $J^{\pm}=J^{p}=J^{z}=U/4$ and anisotropic case with $J^{z}=U/4,J^{\pm}=J^{p}=0$.}
\label{Fig:three}
\end{figure}

In order to reveal underlying physical picture for the OSMT and understand dependence of the OSMT on the spin flip term, we calculate various spin and charge correlations as a function of $U/t$ for both the isotropic and the anisotropic cases. Fig.~\ref{Fig:three} (a) shows that the interorbital charge fluctuations are strongly suppressed both in the anisotropic case as the Mott insulating state appears and in the isotropic case as the OSMT occurs, indicating that decoupling of correlated orbitals do not uniquely leads to the OSMP, in contrast to the existing proposal~\cite{MediciPRB2011} where the OSMT induced by different orbital degeneracies is viewed as a result of orbital decoupling. On the contrary, remarkable difference can be detected between the isotropic and the anisotropic cases in the spin channel. While the squared total spin in $z$ direction, as shown in Fig.~\ref{Fig:three} (b), approaches $1$ as $U/t$ becomes large in the anisotropic case, implying a formation of spin doublet states, i.e., $|S=1,S_z=1\rangle$ and $|S=1,S_z=-1\rangle$, it goes to $2/3$ in the isotropic case as $U/t$ increases, pointing to a fact that spin triplet states due to the spin flip term become dominant in the ground state at large $U/t$.

\begin{figure}[htbp]
\includegraphics[width=0.48\textwidth]{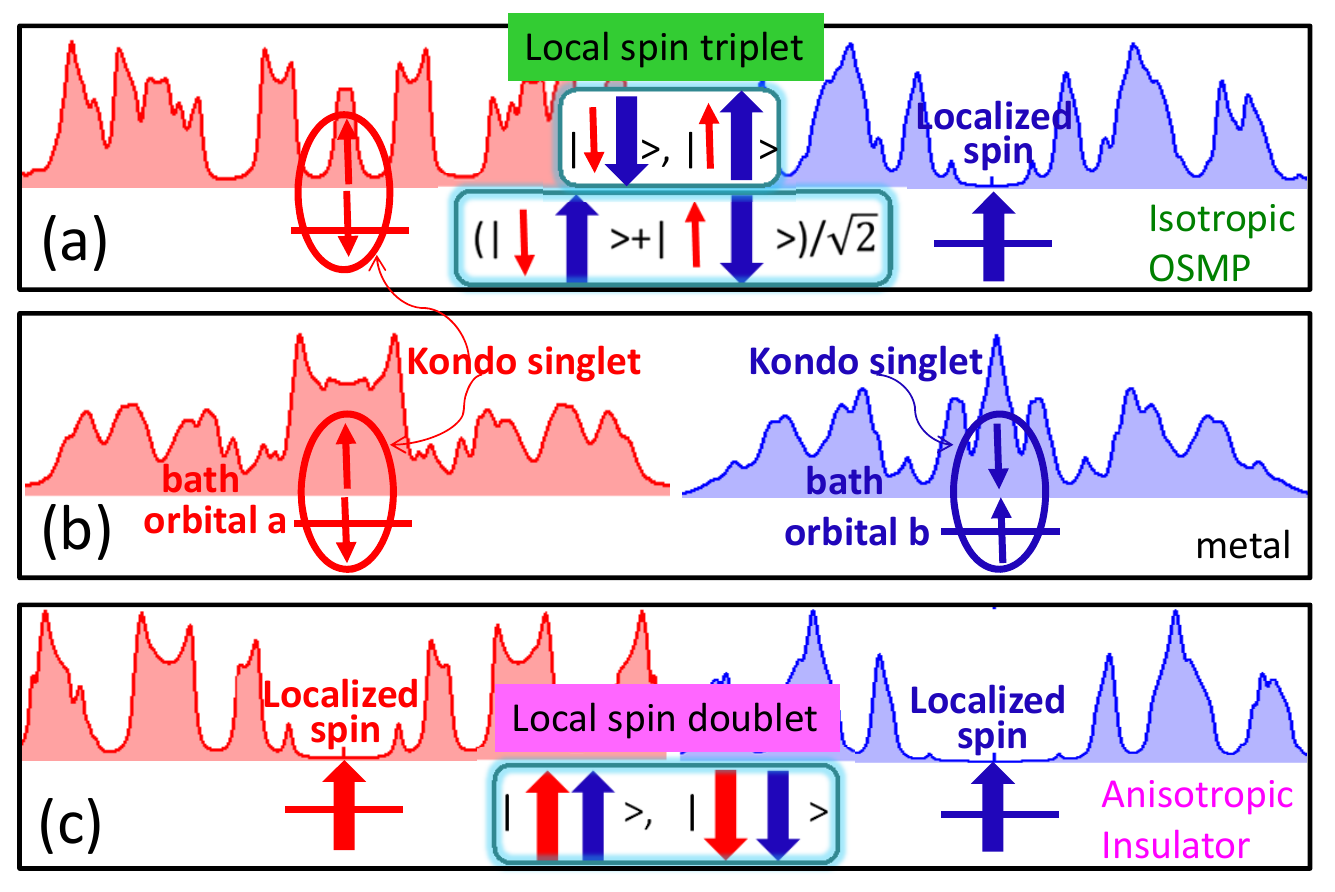}
\caption{(Color online) Cartoon for understanding the importance of spin-flip term on the OSMT, based on a two impurity Anderson model mapped from model~(\ref{eq:hamiltonian}) by neglecting nonlocal correlations. (a), (b), and (c) denote the OSMP in the isotropic case, the metallic state and the Mott phase in the anisotropic case, respectively. As $U/t$ is small (see (b)), both orbitals are strongly hybridized to their own baths, leading to a metallic state. As $U/t$ is larger than a critical value where orbital b is decoupled from its bath and forms a local spin, in the anisotropic case (see (c)), local spin doublet states, induced by the Ising Hund's coupling $J_z$, only allow the electrons with spins parallel to the local spin of orbital b to hop dynamically between orbital a and its bath, leading to a breakdown of the Kondo singlet between orbital a and its bath, and consequently a formation of localized spin also present on orbital a at half filling. On the contrary, in the isotropic case (see (a)), local spin triplet states do not impose any constraint of spin states on orbital a, even if localized spin is formed on orbital b. Therefore, the Kondo singlet remains between orbital a and its bath irrespective of whether the Kondo breakdown happens to orbital b.}
\label{Fig:four}
\end{figure}

The presence of spin triplet states is crucial for the occurrence of the OSMT as illustrated in Fig.~\ref{Fig:four}. Based on the two impurity Anderson model mapped from lattice Hamiltonian~(\ref{eq:hamiltonian}) by neglecting non-local correlations, the metallic state at small $U/t$ is attributed to the strong Kondo screenings in both orbitals as shown in Fig.~\ref{Fig:four} (b). As $U/t$ is larger than a critical value, the Kondo singlet between orbital b and its bath is broken and a local spin forms on orbital b. In the anisotropic case as seen in Fig.~\ref{Fig:four} (c), local spin doublet states impose strong constraint on the spin state of the electrons in orbital a, i.e., only the electrons with spins parallel to the local spin of orbital b are allowed to hop dynamically between orbital a and its bath, while the electrons with antiparallel spins are excluded. As a consequence, the Kondo singlet between orbital a and its bath has to be broken simultaneously as the electron on orbital b is localized. And at half-filling, a Mott insulator appears. However, in the isotropic case (see Fig.~\ref{Fig:four} (a)), additional local $|S=1,S_z=0\rangle$ state, i.e. $\left(|\uparrow_a\downarrow_b\rangle+|\downarrow_a\uparrow_b\rangle\right)/\sqrt{2}$ state, provides a channel for the electrons with antiparallel spins freely hopping onto orbital a irrespective of the spin state in orbital b. Therefore, the OSMP is energetically favored due to the gain of kinetic energy by preserving the Kondo singlet between orbital a and its bath without any cost of potential energy monitoring by the Hund's coupling. The presence of spin triplet states, in some sense, can be viewed as an effective decoupling of the orbitals in spin channel, rather than charge channel. We find that such an OSMT scenario is conceptually identical to the two-stage Kondo effect in a two impurity Kondo problem~\cite{JayaprakashPRL1981}.

\begin{figure}[htbp]
\includegraphics[width=0.48\textwidth]{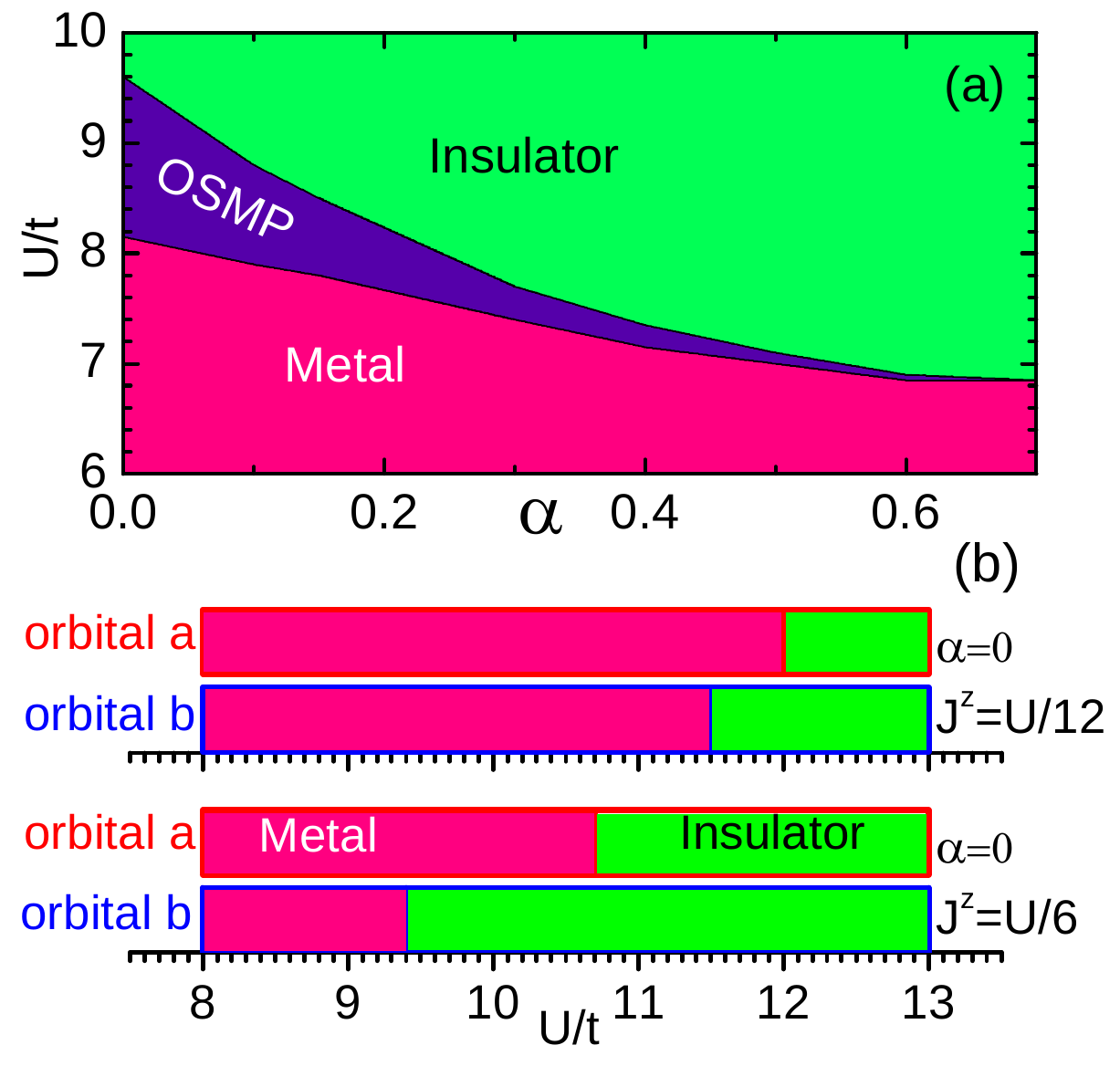}
\caption{(Color online) (a) Phase diagram of the isotropic case with $J^{\pm}=J^{p}=J^{z}=U/4$ in $U/t$-$\alpha$ plane. (b) The Mott transitions in both orbitals at $J^{\pm}=J^{p}=J^{z}=U/12$ and $J^{\pm}=J^{p}=J^{z}=U/6$ as a function of $U/t$ at $\alpha=0$. Here $\alpha=t^{a}_y/t^{a}_x$ and $t^{b}_y/t^{b}_x=1$.}
\label{Fig:five}
\end{figure}

Fig.~\ref{Fig:five} (a) shows a phase diagram in $U/t$-$\alpha$ plane for the isotropic case. It is found that the OSMP induced by each orbital having distinct noninteracting DOS shrinks as $\alpha$ increases. This is reasonable since the larger the $\alpha$, the smaller the difference between the noninteracting DOS of the orbitals. We find that at $\alpha=0.7$, the OSMP disappears.  Fig.~\ref{Fig:five} (b) presents dependence of the OSMT on the Hund's coupling at $\alpha=0$. It is shown that while the OSMP shifts to the region with larger value of $U/t$ as the Hund's coupling becomes smaller, its width is reduced as the Hund's coupling decreases. The OSMP vanishes in the absence of the Hund's coupling due to the presence of strong orbital fluctuations induced by six degenerate two-electron onsite configurations, particularly the spin singlets with both electrons in one of the two orbitals. At $\alpha \neq 0$, the situations are qualitatively the same. Moreover, we replaced the noninteracting DOS of orbital a by the DOS of a honeycomb lattice, and found that the OSMT also occurs. As the OSMP studied in this paper exists in a wide range of model parameters, the proposed origin for the OSMT, i.e., OSMT induced by each orbital with distinct noninteracting DOS, should be realized in nature.

\section{Discussions and Conclusions\label{CD}}

Then, let's discuss the relevance of our results to the materials. The concept of the OSMT is first raised~\cite{AnisimovEPJB2002} in order to understand the metallic state with spin-1/2 local moment observed in Ca$_{2-x}$Sr$_x$RuO$_4$~\cite{NakatsujiPRL2000}. The origin for the OSMT is ascribed to orbitals with unequal bandwidths based on analyzing the band structure of Sr$_2$RuO$_4$, i.e., the bandwidth of $d_{xy}$ orbital is much wider than that of $d_{yz/zx}$ orbital. However, it was found from first principles studies~\cite{FangPRB2004} that the strong RuO$_6$ rotation by Ca substitution reduces the bandwidth of the $d_{xy}$ orbital significantly, but not that of the $d_{yz/zx}$ orbital, which casts doubt on this origin. From our research, it is shown that even if there were no difference in bandwidth between orbitals, the OSMT can still occur due to each orbital with distinct noninteracting DOS. Such a difference in DOS is generally present not only in Ca$_{2-x}$Sr$_x$RuO$_4$~\cite{AnisimovEPJB2002,LiebschPRL2007}, but also in the iron-based superconductors~\cite{MiyakeJPSJ2010}, for example, DOS of $d_{yz/zx}$ orbital resembling a quasi-one-dimensional system (orbital a like) while those of $d_{xy}$, $d_{x^2-y^2}$, and $d_{z^2}$ being more like quasi-two-dimensional systems (orbital b like). In fact, it is the $d_{xy}$ orbital which first encounters a Mott transition in A$_x$Fe$_{2-y}$Se$_2$ (A=K, Rb) superconductors~\cite{YiPRL2013,YuPRL2013}. Moreover, our finding that the renormalization factor, i.e., inverse of the effective mass, of orbital b, is smaller than that of orbital a, is consistent with the surprising observation that the mass enhancement of $d_{xy}$ orbital is larger than that of $d_{yz/zx}$ in Sr$_2$RuO$_4$~\cite{MackenzieRMP2003}.

Finally, we find that VOCl is a layered material with open d shell~\cite{GlawionPRB2009}. And two $3d$ electrons occupy the lowest-lying $d_{x^2-y^2}$ and $d_{xz}$ orbitals which exhibit quasi-one-dimensional and quasi-two-dimensional characters, respectively, and are of similar bandwidth. Moreover, the crystal field splitting between these two orbitals is negligible. And VOCl is classified as a multi-orbital Mott insulator experimentally. Therefore, we propose that VOCl under pressure should be an ideal candidate to verify this new origin for the OSMT.

In conclusion, we find that each orbital having distinct noninteracting DOS is a novel origin for the OSMT at half filling and $T=0$. The spin flip term of the Hund's coupling is indispensable for the presence of the OSMT. The underlying physics can be understood by a two-stage breakdown of the Kondo effect. Further investigations to reveal various effects on the OSMT, such as temperature, interorbital hybridization, doping, nonlocal correlations, number of orbitals, crystal field splitting, are very appealing.

\section{Acknowledgement}\label{Acknowledgement}

This work is supported by National Natural Science Foundation of China (No. 11174219), Program for New Century Excellent Talents in University (NCET-13-0428), Research Fund for the Doctoral Program of Higher Education of China (No. 20110072110044) and the Program for Professor of Special Appointment (Eastern Scholar) at Shanghai Institutions of Higher Learning as well as the Scientific Research Foundation for the Returned Overseas Chinese Scholars, State Education Ministry.

\end{document}